\documentclass[preprint2]{aastex}

\shorttitle{X-ray Sources in Groth Strip}
\shortauthors{Miyaji et al.}
\begin{document}
\title{Multiwavelength Properties of the X-ray Sources in the 
Groth-Westphal Strip Field
\footnote{Based on observations from the XMM-Newton,  an ESA science 
mission with instruments and contributions directly funded by ESA member 
states and the USA (NASA). Also based on data collected 
at Subaru Telescope, which  is operated by the National Astronomical 
Observatory of Japan.}}

\author{Takamitsu Miyaji\altaffilmark{1},
Vicki Sarajedini\altaffilmark{2},
Richard E. Griffiths\altaffilmark{1},
Toru Yamada\altaffilmark{3},
Matthew Schurch\altaffilmark{1,4}, 
David Crist\'obal-Hornillos\altaffilmark{5},
Kentaro Motohara\altaffilmark{6}
}

\altaffiltext{1}{Department of Physics, Carnegie Mellon University,
  Pittsburgh, PA 15213 (miyaji@cmu.edu griffith@seren.phys.cmu.edu)} 
\altaffiltext{2}{Department of Astronomy, University of Florida, \\
  P.O. Box 112055, Gainesville, FL 32611 (vicki@astro.ufl.edu) } 
\altaffiltext{3}{National Astronomical Observatory of Japan, \\
  Mitaka 181-8588, Japan; yamada@optik.mtk.nao.ac.jp} 
\altaffiltext{4}{Department of Physics, U. Bristol, Tyndall Avenue, 
     Bristol BS8 1TL, UK}
\altaffiltext{5}{Instituto de Astrof\'{\i}sica de Canarias, 
     38200 La Laguna, Tenerife, Spain} 
\altaffiltext{6}{Institute of Astronomy, University of Tokyo, 
      Mitaka, Tokyo 181-0015, Japan} 

\begin{abstract}
We summarize the multiwavelength properties of X-ray sources 
detected in the 80 ks XMM-Newton observation of the 
Groth-Westphal Strip, a contiguous strip of 28 HST Wide-Field
Planetary Camera 2 (WFPC2) images. Among the $\approx 150$ 
X-ray sources detected in the XMM-Newton field of view, 23 are within
the WFPC2 fields. Ten spectroscopic redshifts are available
from the Deep Extragalactic Evolutionary Probe (DEEP) and
Canada-France Redshift Survey (CFRS) projects. Four of these show 
broad Mg II emission and can be classified as type 1 AGNs. 
Two of those without any broad lines, nevertheless, have [NeV] 
emission which is an unambiguous signature of AGN activity. One
is a narrow-line Seyfert 1 and the other a type 2 AGN.

As a followup, we have made near-infrared (NIR) spectroscopic 
observations using the OHS/CISCO spectrometer for five of the X-ray 
sources for which we found no indication of an AGN activity in the 
optical spectrum.  We have detected H$\alpha$+[NII] emission 
in four of them.  A broad  H$\alpha$ component 
and/or a large [NII]/H$\alpha$ ratio is seen, suggestive of AGN activity. 
Nineteen sources have been detected in the $K_{\rm s}$ band
and four of these are extremely red objects (EROs; $I_{814}-K_{\rm s}>4$). 

The optical counterparts for the majority of the 
X-ray sources are bulge-dominated. The $I_{814}-K_{\rm s}$ color 
of these bulge-dominated hosts are indeed consistent with evolving 
elliptical galaxies, { while contaminations from star formation/AGN
seems to be present in their $V_{\rm 606}-I_{\rm 814}$ color.}   
Assuming that the known local relations among the bulge luminosity,
central velocity dispersion, and the mass of the central blackhole 
still hold at $z\sim 1$, we compare the AGN luminosity with 
the Eddington luminosity of the central blackhole mass. The AGN bolometric 
luminosity to Eddington luminosity ratio ranges 
from 0.3 to 10\%.

\end{abstract} 
 



\section{Introduction}
 
 Deep X-ray images of a patch of the sky show numerous X-ray sources, 
which mainly consist of a mixture of absorbed and unabsorbed active 
galactic nuclei (AGNs). It is now recognized that these AGNs  
make up the bulk of what has been called the ``X-ray Background''. While 
normal galaxies, whose X-ray emission is probably dominated by the 
integration of X-ray binaries, start to emerge \citep{miy_fluct,horn03} 
at the faintest fluxes, the dominant X-ray source population in the 
deep Chandra and XMM-Newton Surveys comes from AGN activities.  Thus 
multiwavelength studies of these X-ray sources are key to understanding 
the detailed history and physical conditions of the formation and the 
growth of the supermassive blackholes (SMBHs), which are now known 
to reside in the centers of almost all galaxies with a bulge 
\citep{magorrian,merritt}. 


The region known as the ``Groth-Westphal Strip'' (GWS), consists of
Hubble Space Telescope (HST) Wide-Field Planetary Camera 2 (WFPC2) 
medium-deep images of a strip of 28 contiguous fields\citep{groth}.
It is a particularly useful field for extensive multiwavelength 
studies. Numerous on-going and future followup projects have been/
are being conducted on and around this field. Existing morphological 
information from the original WFPC2 observations, combined with the
Deep Extragalactic Evolutionary Probe (DEEP)
\footnote{http://deep.ucolick.org/}, Canada-France Redshift 
Survey (CFRS)\footnote{http://www.astro.utoronto.ca/$\sim$lilly/CFRS/} 
and the ongoing DEEP 2 \footnote{http://deep.berkeley.edu/}
redshift surveys, is providing us with the first clues to the nature of 
the X-ray source counterparts.

Because of the contamination from starlight in the host galaxy, optical
searches for faint AGNs in deep survey fields such as GWS need elaborate efforts. 
Such attempts have been made 
by searching for an unresolved component at the centers of 
galaxies, searching for variable nuclei, and by 
selecting those with ultraviolet-excess cores 
\citep{vicki99,vicki03,beck}.  These surveys reveal 
up to 10\% of galaxies as AGN candidates with nuclei extending as faint as
M$_B$$\simeq$ --15.  
Optical searches, however, are less sensitive to AGNs obscured by dust around
the active nucleus.
On the other hand, X-ray surveys for AGNs are not hindered by the luminosity of the 
underlying host galaxy. In particular, X-ray surveys with hard band ($E>2$ keV) 
sensitivity are also sensitive to the obscured AGNs.

In view of this, we have obtained an 80 ks exposure the northeast part of the GWS
with XMM-Newton.
This field also has the advantage of a low column density of neutral gas 
in our galaxy, corresponding to $N_{\rm H}=1.3\;10^{20}$ cm$^{-2}$ 
\citep{dicky}.  A quick look view and the preliminary Log N-Log S relations 
from these data have been presented in \citet{miy_moriond} and 
\citet{miy_sant}. In this paper, we present the nature of the 23 X-ray 
sources in the GWS, where morphological properties 
from the WFPC2 images and some redshifts are available from the DEEP/CFRS 
redshift surveys. We further make supplemental near infrared spectroscopic 
observation for some of these X-ray sources for which we did not observe 
AGN signatures in the optical spectra. $K_{\rm s}$ band
photometry of the X-ray sources are also presented.

 The scope of this paper is as follows. In Section \ref{sec:xobs},
we describe the X-ray data and analysis, including source detection
and spectral analysis. In Section \ref{sec:opt_ir}, we explain the source 
of the optical and near infrared (NIR) data. The optical and NIR nature of the 
XMM-Newton sources and related statistics are presented.  In \ref{sec:disc}, 
we discuss the overall results on these X-ray sources and black hole mass and the 
relationship with the bulge luminosity. We also comment on selected individual 
sources.  A summary is given in 
\ref{sec:sum}. Throughout this paper, we use 
$H_{\rm 0}=70\;{\rm km\,s^{-1}\, Mpc^{-1}}$, $\Omega_{\rm m}=0.3$,
and  $\Omega_{\rm \Lambda}=0.7$. Unless otherwise noted, $L_{\rm x}$
is the 2-10 keV rest-frame luminosity in units of
${\rm erg\;s^{-1}}$ calculated using the cosmological parameters
shown above.
   
\section{X-ray Data and Analysis}\label{sec:xobs}

The field has been observed with XMM-Newton as part of a
Guaranteed Time program (PI: Griffiths). After cleaning for 
background flares, we have obtained a total of 69 ks
and 81 ks of exposure for EPIC PN and MOS respectively. The 
log of observation is shown in Table \ref{tab:log}.
   

\begin{table}

\caption{Log of XMM-Newton Observations\label{tab:log}}

\footnotesize
\begin{tabular}{cll}
\tableline\tableline
Rev/OBSID & cleaned exp. &Obs. Date \\
\tableline
0113/0127921001 & 51.(PN)/56.(MOS) & 21-Jul.-2000\\
\\
0114/0127921101 & 18.(PN)/25.(MOS) & 23-Jul.-2000\\
 `` /0127921201 &                  &             \\
\tableline
\end{tabular}
\normalsize


\end{table}

 Firstly we have generated an X-ray image for each of the PN, MOS1,
and MOS2 data from XMM-Newton revolutions 0113 and 0114.   
 We have determined the positions of the counterpart X-ray 
sources by Gaussian fits for two broad line AGNs detected in the 
Canada-France Redshift Survey \citep{cfrs3} and a few objects in the Medium Deep 
Survey (MDS) database \citep{mds} with trivial X-ray counterparts.  
Based on the positions of the optical
counterparts, we have determined the linear transformation 
coefficients of the X-ray source positions. During this process,
we did not use X-ray sources which fell in the gaps of the CCDs.
Using the coefficients, we then aligned each cleaned EPIC 
event file. 

 Using the aligned events, we made summed images
in three energy bands, corresponding to photon energies of 
0.5-2, 2-4.5 and 4.5-10 keV.  We also made a 2-10 keV image.
In making the images, we have used PATTERN$\leq$4 (12) for the PN (MOS) 
data and applied the
standard screening criteria. In order to avoid the instrumental
line emission feature which dominates the background in 7.33-9.05 keV, 
the channel energy range corresponding to this energy 
range has been excluded from the PN data. Also exposure
maps have been created for each energy band using the task ``eexpmap''.
The MOS exposure maps in each band have been multiplied by the
ratio of the MOS and PN efficiencies in the given band, so that
the source counts from the summed (PN+2MOSs) image divided by the 
summed exposure map gives count rates which can be converted 
to fluxes using the PN response. The slight shift due to the alignment 
process (at most 2$\arcsec$) is negligible in the usage of the 
exposure map.

 The final source detection has been made to the sum of all
PN and MOS data for the three bands. These three images were 
simultaneously fed into the SAS software 'eboxdetect' 
in the ``local'' mode. For each band, we have created the 
``background map'' by  cubic spline fitting the source excluded 
image (divided by the exposure map) using the sources detected in 
the first step. The next step is to use these ``background'' maps 
to re-run the  'eboxdetect' in the ``map'' mode, i.e., search for the
excess over the background map. The final step is to make
a multi-source maximum-likelihood fit over the background map using
the positions of the sources in the previous step as a starting 
model. This has been made using the SAS task ``emldetect''.
The images in the soft (0.5-2 keV), medium (2-4.5 keV), and 
ultra-hard (4.5-10 keV) bands have been simultaneously fit.  
The procedure also calculates the source detection likelihood
for each band as well as for the sum of these. We accepted
sources with a likelihood parameter of
$$
ML=-\ln (1-P) >14
$$  
where $P$ is the probability that the source exists. We also
made a separate source detection to the 2-10 keV image to 
obtain fluxes of the sources in a more traditional energy
band.
 
 One of the caveats is that cubic spline fits generated using
the SAS procedure ``esplinemap'' sometimes generate spurious 
wavy structures depending on the choice of the number of
spline nodes (the esplinemap parameter {\em nsplinenodes}).
This is probably because some structures in the input image are not 
well represented by the spline function for the given number of nodes.  
In order to obtain the best background map, we have manually 
varied the number of spline nodes until a satisfactory background
map is obtained. While this strategy involves a certain subjectiveness, 
we have checked the goodness of this strategy with extensive simulations. 
The basic results of this analysis are presented in \citet{miy_sant}.
While the source detection should be improved for further quantitative
analysis \citep{baldi,valtchanov}, our current source detection 
scheme provides a sufficiently complete list of X-ray sources for the 
purpose of this paper which is focused on the multiwavelength properties 
of the X-ray sources within a limited area.  These have been verified by visual 
comparisons of the images and the detected 
source lists. The detected source counts have been converted into 
``effective'' count rates by dividing by the value of the exposure map 
smoothed with a $\sigma=12\arcsec$ gaussian.     

 In this paper, we concentrate on the sources which fall within the HST WFPC2 
fields of view. Sixteen of the original 28 GWS fields overlap with the 
XMM field of view.  Additionally, there is another WFPC2 field in this 
general field (PI Lilly). We excluded from this paper a few X-ray sources 
located close to the edge of the WFPC2 field if their optical 
counterparts were not included in the MDS database.  Table \ref{tab:srcs} 
lists the X-ray sources within the WFPC2 FOVs detected using the above 
procedure. In this table, the fluxes have been obtained from the count
rate assuming a power-law or an absorbed power-law spectrum, depending on 
the hardness ratios from detected count rates in the 0.5-2 and 2-10 keV 
bands. If the hardness ratio corresponds to an effective 
photon index of $\Gamma>1.8$, a power-law spectrum is used. For a harder 
spectrum, an absorbed spectrum with fixed $\Gamma=1.8$ and varying 
intrinsic absorption column density ($N_{\rm H}$) placed at $z=1$ 
(a typical redshift of the sources) is used. As long as the value 
of $N_{\rm H}$ is matched to the observed hardness ratio, the count 
rate to flux conversion factor is insensitive to the assumed redshift. 
 

\begin{deluxetable}{ccrrrrrrr}
\tabletypesize{\scriptsize}
\tablecaption{XMM-Newton Sources in the WFPC2-Strip\label{tab:srcs}}
\tablewidth{0pt}
\tablehead{
\colhead{X-no.} & \colhead{Name} & \colhead{$r_{\rm 90}$} & 
\colhead{XMM} & \colhead{$S_{\rm x14}$} & \colhead{$S_{\rm x14}$} 
& \colhead{$ML_{\rm tot}$} & \colhead{$ML_{\rm sft}$} & \colhead {$ML_{\rm hrd}$}\\
& & [$\arcsec$] & counts  & (0.5-2 keV) & (2-10 keV) &\\
}
\startdata
  x8 & XMMGWS J141734.8+522809 & 2.3 & 1000& 2.32$\pm$0.10 &3.03$\pm$0.24& 2581 & 2397 &   340\\
  x9 & XMMGWS J141700.6+521918 & 2.3 & 1083& 1.63$\pm$0.06 &1.16$\pm$0.12& 3005 & 3025 &   224\\
 x10 & XMMGWS J141651.3+522046 & 2.3 & 1082& 1.46$\pm$0.06 &2.17$\pm$0.17& 2844 & 2500 &   463\\
 x11 & XMMGWS J141642.4+521813 & 2.4 &  910& 1.21$\pm$0.06 &4.01$\pm$0.23& 2335 & 1490 &   835\\
 x20 & XMMGWS J141741.7+522822 & 2.4 &  635& 0.93$\pm$0.05 &2.35$\pm$0.19& 1463 & 1087 &   371\\
 x22 & XMMGWS J141653.2+522104 & 2.4 &  681& 0.80$\pm$0.04 &1.82$\pm$0.16& 1286 &  962 &   326\\
 x26 & XMMGWS J141715.0+522311 & 2.4 &  443& 0.47$\pm$0.03 &0.82$\pm$0.10&  810 &  681 &   132\\
 x28 & XMMGWS J141704.1+522139 & 2.4 &  568& 0.29$\pm$0.03 &3.45$\pm$0.20& 1049 &  202 &   896\\
 x33 & XMMGWS J141816.3+522940 & 2.6 &  409& 1.04$\pm$0.07 &1.06$\pm$0.22&  631 &  614 &    30\\
 x40 & XMMGWS J141653.8+522042 & 2.7 &  187& 0.27$\pm$0.03 &    $<$0.4   &  134 &  151 &     6\\
 x44 & XMMGWS J141729.9+522748 & 2.6 &  241& 0.31$\pm$0.03 &0.72$\pm$0.12&  307 &  224 &    65\\
 x46 & XMMGWS J141745.6+522801 & 2.7 &  204& 0.35$\pm$0.04 &0.58$\pm$0.14&  245 &  233 &    28\\
 x52 & XMMGWS J141746.0+523031 & 2.8 &  179& 0.16$\pm$0.03 &2.01$\pm$0.23&  189 &   44 &   177\\
 x55 & XMMGWS J141633.5+521502 & 3.1 &  152& 0.32$\pm$0.04 &1.21$\pm$0.23&  160 &   93 &    44\\
 x61 & XMMGWS J141758.7+523138 & 3.1 &  135& 0.32$\pm$0.04 &0.94$\pm$0.19&  134 &  108 &    39\\
 x66 & XMMGWS J141649.8+521810 & 3.0 &  112& 0.06$\pm$0.02 &1.15$\pm$0.16&  107 &   10 &   102\\
 x69 & XMMGWS J141749.1+522811 & 3.1 &  129& 0.23$\pm$0.03 &0.41$\pm$0.13&  121 &  111 &    13\\
 x83 & XMMGWS J141639.7+521502 & 4.3 &   73& 0.16$\pm$0.03 & $<$0.7     &   33 &   31 &     2\\
x118 & XMMGWS J141753.1+522840 & 3.8 &   73& 0.13$\pm$0.03 & $<$0.7     &   47 &   41 &     6\\
x125 & XMMGWS J141756.9+523125 & 4.2 &   68& 0.16$\pm$0.03 &0.54$\pm$0.17&   44 &   35 &    13\\
x130 & XMMGWS J141754.3+523125 & 4.8 &   52& 0.10$\pm$0.03 &0.51$\pm$0.17&   22 &   14 &    12\\
x133 & XMMGWS J141720.0+522459 & 4.1 &   53& 0.03$\pm$0.02 &0.33$\pm$0.08&   23 &    3 &    23\\
x146 & XMMGWS J141745.8+522950 & 3.9 &   61& $<$0.02 &1.13$\pm$0.21&   31 &    0 &    53\\
\enddata			

\end{deluxetable}

The columns of Table \ref{tab:srcs} are explained below:
\begin{description}
\item[X-No.:] The internal XMM source ID (hereafter referred to as
the X-number), which is used to identify the XMM-Newton sources 
throughout this paper (with or without the preceding ``x'').
\item[Name:] The source name according to the IAU registry in the
  format XMMGWS JHHMMSS.s+DDMMSS.  This also gives the detected 
  XMM source position. Note that the last digits are truncated.
  \footnote{See http://cdsweb.u-strasbg.fr/iau-spec.html)}.
\item[Pos. err:] Positional error (90\% encircled) in arcseconds. 
    See below.
\item[XMM counts:] The detected counts in the total (0.5-10 keV) band.
\item[${\bf S_{\rm x14}(0.5-2)}$:] The soft (0.5-2 keV) band flux, in units
  of $10^{-14}{\rm erg\,s^{-1}\,cm^{-2}}$. 
\item[${\bf S_{\rm x14}(2-10)}$:] The hard (2-10 keV) band flux, in the same  units.
\item[${\bf ML_{\rm tot}}$:] The source existence likelihood (see above) in the total band.
\item[${\bf ML_{\rm sft}}$:] The source existence likelihood in the soft band.
\item[${\bf ML_{\rm hrd}}$:] The source existence likelihood in the hard band.
\end{description}

While we treated sources with $ML>14$ for the total band as detected and
included them in Table \ref{tab:srcs}, in each of the soft and hard bands, 
the $ML$ values are smaller.  In the soft and hard bands, we list 
a nominal flux and 1$\sigma$ error if $ML\geq 10$ and 
a 3$\sigma$ upper limit for a lower significance. 


\begin{figure*}
\epsscale{1.8}
\plotone{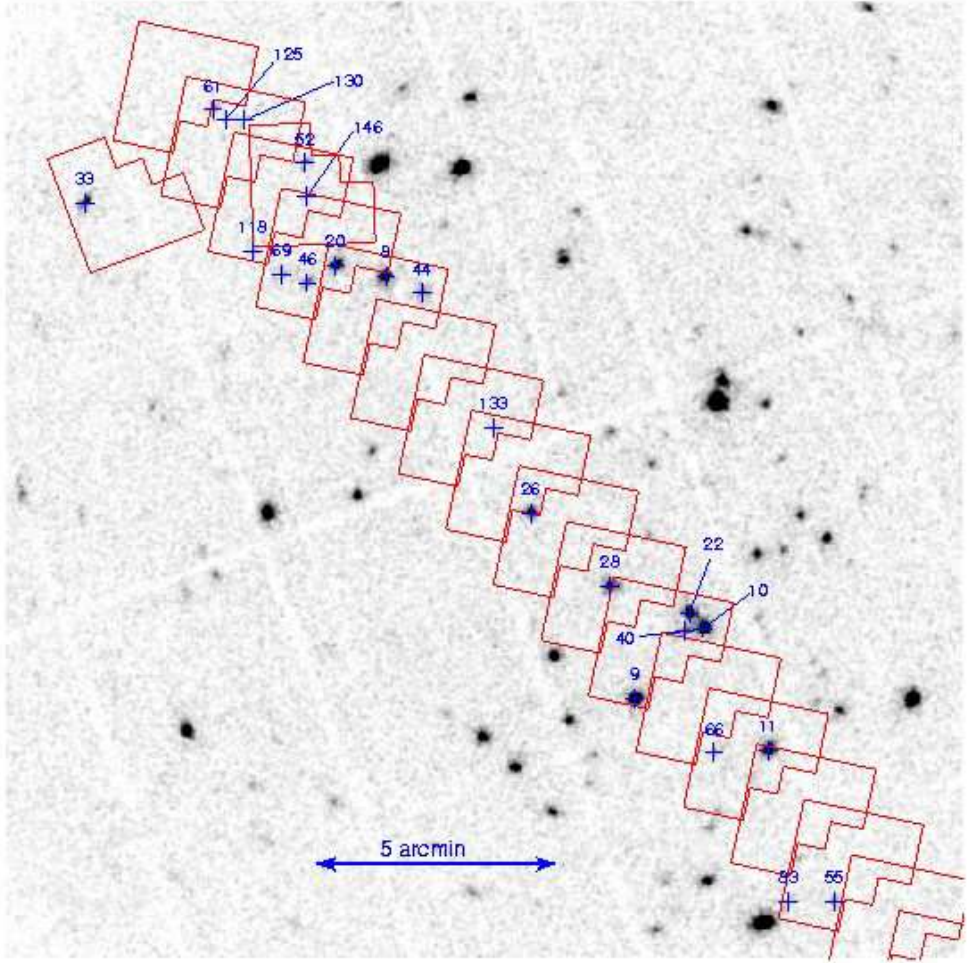}
\caption
 {The WFPC2 fields of view are overlaid on the full band 
(0.5-10 keV) XMM-Newton image of the GWS. The underlying
image has been smoothed with a $\sigma=2\arcsec$ Gaussian. The positions 
of the X-ray sources in Table \ref{tab:srcs} are marked.
}
\label{fig:xmmfield}
\end{figure*}

 Fig. \ref{fig:xmmfield} shows the summed full-band (0.5-10 keV) 
XMM-Newton image smoothed with a $\sigma=2\arcsec$ Gaussian for display. 
The HST WFPC2 FOVs are overlaid and the positions of the 
sources in Table \ref{tab:srcs} are marked with crosses with 
the X-number designations.

 There is a 200 ks Chandra ACIS-I observation (PI Nandra) covering 
the northeastern part of the field and the data came to public archive 
immediately after the observation.  We have also generated a source 
list from the Chandra data. We have aligned the Chandra 
events in the same way as the XMM-Newton data.  Within 7.4 arcminutes 
of the Chandra pointing center, the Chandra data provide much more 
accurate X-ray source positions and we used them for unambiguous 
identification of the X-ray source counterparts as shown below. In 
this paper, we choose not to discuss X-ray sources detected in the 
Chandra data but not detected in the XMM-Newton data, because they are  
fainter than those in this paper and the current optical 
data and redshift information are very limited.  

 The uncertainties of the source positions have been determined
by the combination of the empirically determined systematic errors
and the statistical errors. The former has been determined by
the rms residuals of the alignment procedure described above. These
are $\sigma_{\rm sys}=1\farcs 5$ for the XMM data and $0\farcs 7$ for 
the Chandra data along each of the x and y axes.  
For fainter sources and/or sources with large off-axis angles, 
the statistical error of the source position ($\sigma_{\rm stat}$) 
becomes comparable to $\sigma_{\rm sys}$. Since the $\sigma_{\rm stat}$ 
is much smaller than $\sigma_{\rm sys}$ for brighter sources, including 
those used for the alignment process,  $\sigma_{\rm stat}$ and $\sigma_{\rm sys}$ 
are independent of each other. Thus we can estimate the 90\% encircle radius 
($r_{\rm 0.9}$) by
\begin{equation}
r_{\rm 0.9}= 2.15\;[\sigma_{\rm sys}^2+\sigma_{\rm stat}^2],
\end{equation}
where 2.15 is the conversion factor between 1$\sigma$ and 90\% encircle
radius assuming a two-dimensional Gaussian probability distribution.   

 The XMM X-ray sources with improved positions from the Chandra
data are summarized in Table \ref{tab:chanpos}.


\begin{table}
\caption{Chandra Positions\label{tab:chanpos}}

\begin{tabular}{clll}
\tableline\tableline
  XMM  &  \multicolumn{3}{c}{Chandra Position (J2000)}\\
{X-no.} & RA($\degr$) & DEC($\degr$) &
 $r_{\rm 90} {\rm (\arcsec)}$\\
\tableline
  x8 & 214.39514 & 52.46958 & 0.7\\ 
 x20 & 214.42441 & 52.47316 & 0.7\\ 
 x26 & 214.31275 & 52.38685 & 0.8\\ 
 x44 & 214.37476 & 52.46324 & 0.7\\ 
 x46 & 214.44027 & 52.46722 & 0.7\\ 
 x52 & 214.44140 & 52.50901 & 0.7\\ 
 x61 & 214.49547 & 52.52754 & 0.8\\ 
 x69 & 214.45498 & 52.46987 & 0.8\\ 
x125 & 214.48674 & 52.52354 & 0.8\\ 
x130 & 214.47583 & 52.52329 & 0.8\\ 
x133 & 214.33364 & 52.41676 & 0.8\\ 
x146 & 214.43923 & 52.49754 & 0.8\\ 
\tableline
\end{tabular}
\end{table}

\subsection{X-ray Spectral Properties}

\begin{figure}
\epsscale{1.0}
\plotone{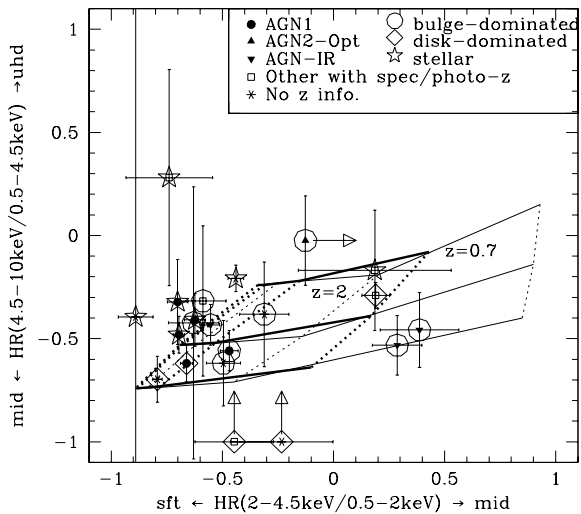}
\caption
  {The X-ray color-color diagram of the X-ray sources in GWS. The inner
symbol shows the spectral classification (see Sect. \ref{sec:opt_ir}; one
narrow-line Seyfert 1 has been included among AGN1s) and the outer symbol shows 
the morphological classifications as labeled. Error bars show 1$\sigma$ errors. 
The grid shows the location for absorbed power-law spectra at z=0.7 
(wider grid) and z=2 (narrower grid, shown in thick lines). The solid 
lines correspond to photon
indices of $\Gamma=$ 1,2, and 3 from top to bottom and the dotted lines
to ${\rm Log}\; N_{\rm H}=-\infty,21.5,22.5,$ and 23.5 cm$^{-2}$ from left 
to right respectively. Objects x44 (X-ray source is off-nucleus) and x83 (Galactic
star) are excluded from the plot.
}
\label{fig:hr}
\end{figure}

 As a preview of the X-ray spectral properties of the detected
sources, we plot a hardness ratio diagram in Fig. \ref{fig:hr}.
The hardness ratio between the 2-4.5 keV and 4.5-10 keV is
plotted against that between 0.5-2 keV and 2-4.5 keV. The
typical 1$\sigma$ errors of the hardness ratios are 0.05--0.2;
errors exceeding 0.4 are shown.  We overplot grids 
showing the loci of absorbed power-law spectra for varying
values of the photon index $\Gamma$ and intrinsic absorption
$N_{\rm H}^{\rm z}$ for z=0.7 and z=2 respectively as described in
the figure caption (Galactic $N_{\rm H}=1.3\; 10^{20}$ cm$^{-2}$ 
has been included). Key optical/IR properties (Sect. \ref{sec:opt_ir})
of  the X-ray sources are indicated. 

\placefigure{fig:hr}

For bright sources, (XMM counts $\gtrsim$ 300), X-ray pulse-height 
spectral analysis gives further constraints on the spectral parameters.
 For the spectral analysis, we have used the
data from the first observation in Table \ref{tab:log}
(Rev/OBSID = 0113/0127921001), which contain mostly good intervals
free from background flares. The extraction of the spectra has 
been made using the SAS analysis package V5.4 and we have 
used the standard screening criteria with
PATTERN$<5$ and PATTERN$<13$ for PN and MOS respectively as
before. The calibration files used to generate the calibrated data 
products and response files for the spectral analysis (using 
the SAS task ``epproc'' and ``emproc'') were up to date as 
of Aug. 7, 2003. We extracted the source spectra using a 
circular region with a typical radius of 25$\arcsec$, depending on the source 
count.  The background spectra have been extracted using an annular
region surrounding the source with inner and outer radii of 
30$\arcsec$ and 60$\arcsec$ respectively. We have further jointly
analyzed two co-added MOS spectra and the PN data.

 For sources with $>600$ counts, we could reasonably constrain the
parameters with a $\chi^2$ fit to the background subtracted spectra 
rebinned to have at least 20 counts per bin. For fainter sources, we have
utilized the XSPEC implementation of the C-statistics \citep{cash}
to the spectra with much finer binnings. Because the C-statistics cannot be 
applied to background-subracted data, the background and (source+background)
spectra have been jointly fit. The background has been approximated by
one or two power-laws and a broad Gaussian.  This background model
was then scaled by the source and background extraction areas and added 
to the source model.  The result was then fitted to
the source spectrum without background subtraction.

\placetable{tab:spec}

\begin{deluxetable}{ccccccc}
\tabletypesize{\footnotesize}
\tablecaption{Results of the X-ray Spectral Analysis\label{tab:spec}}
\tablewidth{0pt}
\tablehead{
\colhead{X-No.} & \colhead{$z$\tablenotemark{a}} & \colhead{$\Gamma$} &  
\colhead{$N_{\rm H}^{\rm z}$} & \colhead{$S_{\rm x14}$\tablenotemark{b}} & 
\colhead{${\rm Log}\;L_{\rm x}$\tablenotemark{c}} & \colhead{$\chi^2/\nu$\tablenotemark{d}}\\
& & &  
\colhead{$10^{22}{\rm cm^{-2}}$} & \colhead{${\rm erg\,s^{-1}\,cm^{-2}}$} & 
 \colhead{${\rm erg\,s^{-1}}$} & \colhead{or Prob.(\%)}\\
}
\startdata
x8  & 1.223 & 2.2(2.0;2.5) & $<$1.3     &  1.6 &  44.2 &  22./38\\
x9  & 0.5*  & 2.5(2.3;2.9) & $<$0.07    &  1.0 &   ... &  20./47\\
x10 & 0.801 & 2.2(2.0;2.8) & $<$0.3     &  1.4 &  43.7 &  7./29\\
x11 & 1*   & 1.7(1.4;2.2) &1.6(0.7;2.5) &  3.5 &   ... &  16./31\\
x20 & 1.148 & 2.0(1.6;2.7) &1.8(0.7;3.3)&  1.9 &  44.2 &  9./19\\
x22 & 0.983 & 1.4(1.2;1.7) &  $<$0.5    &  2.7 &  44.2 &  11./20\\
x28 & 0.76p & 1.2(0.8;1.7) &3.0(1.7;5.3)&  4.1 &  43.8 &   62.  \\   
\enddata
\tablenotetext{a}{Objects without spectroscopic redshifts are marked with an asterisk and
the assumed redshift for $N_{\rm H}$ calculation is shown. Photometric redshifts 
are marked with ``p''.}
\tablenotetext{b}{Source flux in the 2-10 keV band at the observer's frame,
 corrected for Galactic absorption.}
\tablenotetext{c}{The base 10 logarithm of the source luminosity in 
the 2-10 keV band at the source rest frame corrected for 
Galactic and intrinsic absorption.} 
\tablenotetext{d}{For the $\chi^2$ fitting results, $\chi^2$ and the degrees 
 of freedom are shown. For fits using C-statistics, we give the percentage of 
 simulated spectra drawn from a Gaussian distribution centered on the best 
 fit with sigma from the covariance matrix which gives C-values smaller than 
 the observed. (Using the xspec command ``goodness'' with the 
 ``sim'' option.)} 
\end{deluxetable}

The source spectrum is fitted with a power-law, Galactic
absorption (fixed at $N_{\rm H}=1.3\;10^{20}$ cm$^{-2}$), and an intrinsic 
absorption at the source redshift. For the sources without known 
redshifts, we have assumed a redshift for determining 
the value of $N_{\rm H}^{\rm z}$. The results have been summarized 
in Table \ref{tab:spec}. See comments to this table for details.

\section{Optical Identifications and NIR Observations}
\label{sec:opt_ir}

This field has been targeted by a number of past and future
survey projects in many wavelength, starting with the 28 contiguous 
HST WFPC observations \citep{groth}.  The main
source of the optical data summarized in this paper comes from
the HST MDS \citep{mds} database and the DEEP survey.
We have also made supplemental NIR spectroscopic observations 
for several of the X-ray sources. Infrared ($K_{\rm s}$)
photometry of these sources from the EMIR-COSMOS project\citep{Cristobal03} 
are also included.  The basic results are summarized in 
Table \ref{tab:opt-ir}.

\begin{deluxetable}{cccccccccccccc}
\tabletypesize{\scriptsize}
\rotate
\tablecaption{Optical-Infrared Morphological and Spectroscopic Properties
\label{tab:opt-ir}}
\tablewidth{0pt}
\tablehead{
\colhead{(1)} & \colhead{(2)} & \colhead{(3)} & \colhead{(4)} 
& \colhead{(5)} & \colhead{(6)} & \colhead{(7)} & 
\colhead{(8)} & \colhead{(9)} & \colhead{(10)} & 
\colhead{(11)} & \colhead{(12)} &  \colhead{(13)} & \colhead{(14)}\\

\colhead{X-No.} & \colhead{MDS-ID} & \colhead{$V_{\rm 606}$\tablenotemark{a}} & 
\colhead{$I_{\rm 814}$\tablenotemark{b}}  
& \colhead{B/T\tablenotemark{c}} & \colhead{DEEP-ID} & \colhead{$K_{\rm S}$\tablenotemark{d}} & \colhead{$z$} & 
\colhead{$Q_{\rm z}$} & \colhead{ref\tablenotemark{e}} & \colhead{Class} & 
\colhead{Opt. Line\tablenotemark{f}} & \colhead{IR lines\tablenotemark{g}} 
&\colhead{${\rm Log} L_{\rm x}$}\\
}
\startdata
x8 & u26x7:0004 & 20.6 & 20.2 & S,S  & 083\_5407 & 17.4 & 1.22 & 3 
& D & AGN1 &{\bf MgII} & ... & 44.1\\
x9 & u26xd:0007 & 20.4 & 19.4 & 0.28,0.24 & 144\_2774 & 16.7 
& ... & 0 & ... & ... & ... & ... & ...\\
x10 & u26xd:0011 & 21.4 & 19.8 & S,0.25 & 142\_4838 & 16.9 
 & 0.808 & 5 & D & NLS1 & [NeV] & ...& 43.8\\
x11 & u26xf:0001 & 18.1 & 17.7 & S,S  & 173\_7369 & 15.9
 & ... & 0 & ... & ... & ... & ...\\
x20 & u2ay1:0025 & 23.4 & 21.5 & B,B & 083\_5273 & 16.5
 & 1.148 & 3 & D & AGN1 & {\bf MgII} & ...& 44.2\\
x22 & u26xd:0034 & 22.7 & 21.5 & B,B & 142\_2752 & 18.3 
 & 0.983 & 3 & O & AGN-IR & ... & {\bf H$\alpha$},[NII] & 43.9\\
x26 & u26xb:0010 & 21.2 & 20.9 & S,S  & 123\_2458 & 19.4
 & 1.263 & 3 & D & AGN1 & {\bf MgII} & ... & 43.8\\
x28 & u26xc:0019 & 22.8 & 20.9 & B,B & 133\_1016 & 17.6
 & 0.757 & -1 & I & ... & ... & ... & 44.0\\
x33 & u2iy4:0004 & ... & 19.5 & ...,S  & None & ...
 & 1.603 & 4 & C & AGN1 & {\bf MgII} & ... & 44.3\\
x40 & u26xd:0062? & 23.1 & 23.3 & S,S  & 142\_2530 & $>$21.5
 & ... & 0 & ... & ... & ... & ... & ...\\
x44 & u26x7:0120 & 25.3 & 23.7 & G,B & 082\_5240 & 19.6
 & ... & 0 & ... & ... & ... & ... & ...\\
x46 & u2ay1:0019 & 21.4 & 20.7 & (D,D) & 074\_2638 & 18.9
 & 0.432 & 4 & D & ... & LEX & ... & 42.6\\
x52 & u26x6:0011 & 22.9 & 21.9 & B,0.41 & 062\_2060 & 17.5 
 & 0.985 & 4 & D & AGN-IR & LEX & {\bf H$\alpha$},[NII] & 44.0\\
x55 & u26xh:0177 & 24.7 & 23.7 & D,B & 184\_2148 & $>$20.9  
 & 0.0000 & 0 & ... & ... & ... & ... & ...\\
x61 & u26x4:0021 & 23.0 & 21.6 & 0.51,B & 053\_4446 & 18.5 
 & 0.637 & 1 & C & ... & LEX & ... & 43.2\\
x66 & u26xf:0024 & 22.4 & 20.1 & 0.22,B & 164\_6109 & 17.6
 & 0.808 & 3 & D & AGN-IR & LEX & {\bf H$\alpha$},[NII] & 43.6\\
x69 & u2ay1:0092 & 24.6 & 22.8 & B,B & 074\_6236 & 18.7
 & 0.995 & 3 & O & AGN-IR & ... & H$\alpha$,[NII] & 43.3\\
x83 & u26xh:0001 & 14.8 & 15.6 & S,B  & 184\_7960 & 13.4
 & 0.0000 & 0 & ... & Star & ... & ... & ...\\
x118 & u26x6:0004 & 20.6 & 20.4 & 0.26,0.30 & 064\_2873 & 17.6
 & 0.719 & -1 & B & ... & ... & ... & 42.8\\
x125 & u26x5:0056 & 24.0 & 22.7 & S,S  & 053\_3424 &  19.3
 & 1.554 & -1 & B & ... & ... & ... & 43.9\\
x130 & u26x5:0160 & 25.2 & 24.8 & G,D & 052\_0940 & $>$21.0
 & 0.0000 & 0 & ... & ... & ... & ...& ...\\
x133 & u26x9:0067 & 24.1 & 22.8 & S,S & 113\_6354 & 19.9
 & 0.704 & -1 & B & ... & ... & ...& 42.9 \\
x146 & u26x6:0010 & 22.8 & 21.0 & 0.12,0.21 & 073\_7749 & 17.9
 & 0.873 & 3 & D & AGN2-Opt & [NeV] & ... & 43.8\\
\enddata

\tablenotetext{a}{Errors are smaller than 0.1 except
(errors in parenthesis) x44 (0.1), x55 (0.2) and x130 (0.5).}
\tablenotetext{b}{Errors are smaller than 0.1 except for 
x55 (0.2) and x130 (0.2).}
\tablenotetext{c}{Bulge to total ratio if applicable, or the 
morphological class (B:pure bulge, D:pure disk, S:stellar, G:galaxy) 
for the $V_{\rm 606}$ band followed by that of the $I_{\rm 814}$ band.}
\tablenotetext{d}{Errors are smaller than 0.1 except for 
x26 (0.1), x44 (0.2), x125 (0.2) and x133 (0.1). Lower limit magnitudes
correspond to the 3$\sigma$ detection limit.} 
\tablenotetext{e}{Redshift reference--D: DEEP C:CFRS 
B: \citet{brunner} I: \citet{im} O: Subaru OHS (this work)}
\tablenotetext{f}{Only key features are shown. Broad lines are indicated by 
{\bf boldface}. Those with only low-excitation narrow emission lines 
are labeled as LEX.}
\tablenotetext{g}{Broad lines are indicated by {\bf boldface}.}
\end{deluxetable}


\subsection{Optical Morphology}


\begin{figure}
\epsscale{1.0}
\plotone{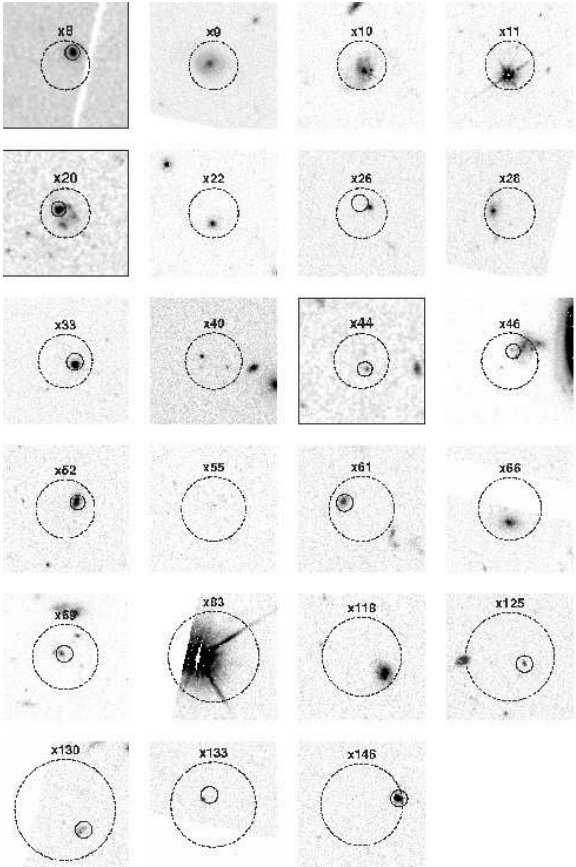}
\caption
 {Postage stamp images from the HST WFPC2 images of the XMM-Newton 
sources in Table \ref{tab:srcs} are shown. Each image has a dimension
of $12\arcsec \times 12\arcsec$.  North is up and east is left. Large 
dashed circles and small solid circles are the 90\% error circles
of XMM-Newton and Chandra sources (if they exist) respectively.
}
\label{fig:poststamp}
\end{figure}

 Fig. \ref{fig:poststamp} shows the HST WFPC2 F814W images (with the exception 
of x8, x20, and x44, where the F606W images are shown) with $r_{\rm 0.9}$ 
error circles for XMM and Chandra (where applicable) source positions. 
Because each WFPC2 image may have absolute astrometric error of 
$\sim 1\arcsec$, in a few cases, the X-ray sources in one WFPC2 FOV 
show systematic offsets from their counterparts. In those cases, we have 
shifted the WFPC2 image by matching the positions of one of the X-ray 
sources with its unambiguous counterpart in making overlays.
In Fig. \ref{fig:poststamp}, the overlays for x8, x20, and x44 have 
been tuned using x8, those for x61, x125, and x130 using x61, those for
x52, x146, and x118 using x146, and those for x69, x46 and x20 using x69. 
In any case, a translational shift without a rotation gives a satisfactory 
correction, where all the unambiguous counterparts (including Chandra sources
without XMM detections) are within the Chandra error circles. 
Note that the drawn error circles ($r_{\rm 0.9}$) may be an overestimate in 
these cases.       

 All of the X-ray source counterparts have an entry in the MDS 
database\footnote{http://archive.stsci.edu/mds}, where
the results of the disk-bulge decomposition are provided. The entries from
the MDS database, including the MDS-ID, magnitudes in F814W (or $I_{\rm 814}$) 
and F606W (or $V_{\rm 606}$) filters, and the bulge-to-total brightness ratios 
(B/T) (or other morphological classifications) have been shown in Table \ref{tab:opt-ir} 
in columns (2),(3),(4), 
and (5) respectively. The first and second entries in column (5) 
(separated by a comma) are for the F606W and F814W bands respectively.
If an entry in column (5) shows a number between 0 and 1, it means that 
the host galaxy has been deconvolved with an exponential disk+bulge model 
and the bulge-to-total light ratio is shown. Other possible entries in 
this column are ``B'' (pure bulge), ``D'' (pure disk), ``S'' (unresolved 
point source), and  ``G'' (galaxy; emission is extended, but could not 
discriminate between the exponential disk and bulge models).  

\subsection{Optical Spectroscopic Classification}

 Optical spectra are available for 8 of our X-ray sources
through the DEEP survey\footnote{The spectra obtained by the 
DEEP team are publicly available and in the DEEP 
Public Database (http://saci.ucolick.org/verdi/public/index.html).
Thus they have not been reproduced here.}
and two from the CFRS. Key AGN classification emission lines
are shown in column (12) of Table \ref{tab:opt-ir}. Broad permitted lines 
are indicated in boldface. 

In summary, four X-ray sources show broad Mg II 
emission lines and we classify these as type 1 Seyfert/QSOs. They are identified with 
``AGN1'' in column (12) of Table \ref{tab:opt-ir}. Other objects with
optical spectra show only narrow permitted and/or forbidden lines. 
Two of these show [NeV] emission, which is an unambiguous indicator of 
an AGN activity. One of these (x10) is a narrow-line Seyfert 1 galaxy 
(``NLS1'' in column [11]) as discussed in Sect. \ref{sec:indi}.  
We tentatively classify the other object (x146) as a type 2 AGN and 
classify it as  ``AGN2-Opt''. 

The other four objects do not show signs of AGN activity in their 
optical spectra in that they show neither  broad permitted lines nor
high excitation lines.  These are classified
as ``LEX'' (low-excitation emission line objects) in column (12).

\subsection{Subaru CISCO/OHS NIR Spectroscopy}

 We have made NIR spectroscopic observations for a subset
of the X-ray sources { with no optical spectroscopic
indication of an AGN activity. These include objects with optical spectra
which show only low excitation lines as well as those without optical 
spectra.}
Observations were made using the CISCO spectrometer
with the OH suppressor (CISCO/OHS) \citep{motohara,iwamuro} 
attached to the Nasmyth focus of the 8.2m Subaru Telescope on March 22-24, 
2003. The JH-Grism with a $0\farcs 5$ slit width was used. The spectral resolution 
of this configuration is $\lambda/\Delta \lambda = 400$. 

 The primary purpose of the NIR spectroscopy was to detect the  
redshifted H$\alpha \lambda 6563$+[NII] $\lambda\lambda 6548,6583$ 
complex and investigate the AGN nature of these X-ray sources.  
{ The ratio of the strengths of the [NII] doublet and H$\alpha$ 
is a good discriminator between the AGN and starburst origins
of the ionizing radiation.} Typical Seyfert 2 galaxies have 
${\rm log ([NII]\lambda 6583/H\alpha) >}$-0.25
\citep{veillux}. 
{ We also intended to search for broad components in  H$\alpha$.
Since a dust cloud is more transparent to redder light, 
moderately obscured AGN activities are much more visible in the 
longer wavelength lines such as H$\alpha$
rather than shorter wavelength lines such as H$\beta$ or Mg II.} 
In fact, there have been a number of reported cases  
where X-ray source counterparts show broad H$\alpha$ lines, even though they 
only show narrow lines in the shorter wavelength optical spectra 
\citep{akiyama_q2,schm98}.  

{ We did not include those with z$<$0.7 (x46, x61) in our
NIR spectroscopy targets, because the H$\alpha$+[NII] line complex 
does not fall within the wavelength range observable by CISCO/OHS JH-grism. 
Also in order to make spectroscopic observation feasible within a reasonable 
exposure time, we limited our spectroscopic observation to those with an estimated 
H-band magitude of $H<21$ \citep{iwamuro}, where H band magnitudes were 
estimated by $H \approx I_{\rm 814}-1.4$ \citep{barger01}. Five from the 
23 XMM-Newton sources were observed during our observing run.}  
  
\begin{figure}
\epsscale{1.0}
\plotone{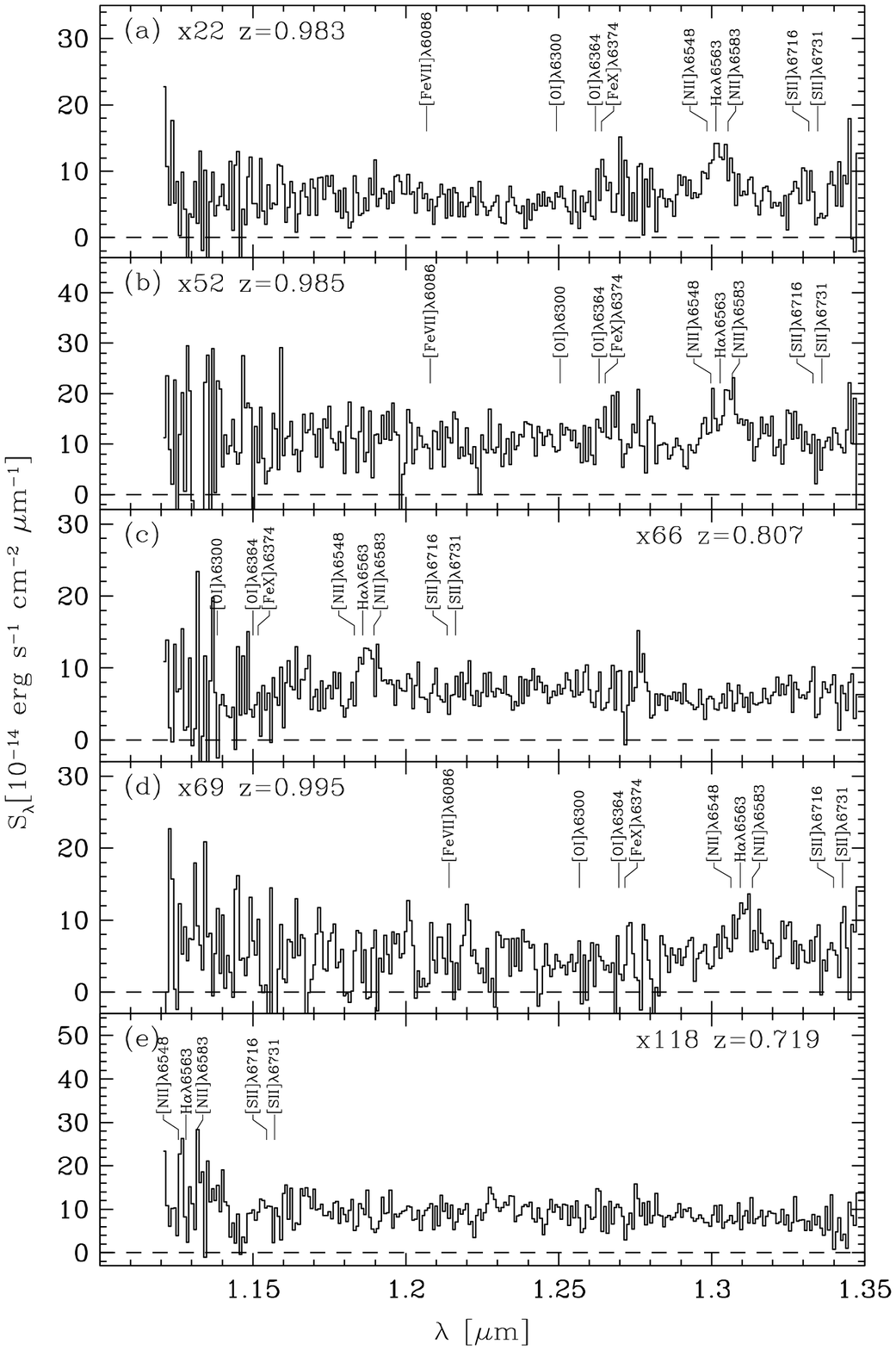}
\caption
 {Near-infrared (NIR) spectra of the XMM-Newton sources observed with
 CISCO-OHS on the Subaru Telescope. Redshifted positions
 of common  QSO emission lines from \cite{wilkes_quant} are marked.
 }
\label{fig:ohs}
\end{figure}

The resulting spectra, reduced following the procedure by \citet{iwamuro}, 
are shown in Fig \ref{fig:ohs}. Preliminary flux calibrations have been made 
using a nearby standard star observed immediately after the target.


We have detected the H$\alpha$+[NII] complex in 4 of the 5 observed 
X-ray sources.
In all four cases, the spectra suggest a possible broad 
H$\alpha$ component (FWHM $\sim 2000 {\rm km\,s^{-1}}$) and/or 
strong [NII] lines. This is indicated in column (13) of 
Table \ref{tab:opt-ir} and the sources are classified as  ``AGN-IR'' 
in column (11).  

Because of the weak signal-to-noise ratio and difficulty in determining 
the background, careful analysis is required to make any 
quantitative conclusions. This will be the topic of a separate paper that 
will describe the details of the observations, data reduction, and 
quantitative analyses for the Subaru CISCO-OHS spectra of the 
sources shown in this paper and a few additional sources observed
during the observation run.

\subsection{Infrared ($K_{\rm s}$) Photometry}

 The $K_{\rm s}$ photometry was provided by the EMIR-COSMOS project 
 \citep{Guzman03,Balcells03}. The GWS was 
imaged with the near-infrared camera INGRID on the William Herschel 
Telescope in the Spanish Observatorio del Roque de los Muchachos during 
two runs in 2000 April and 2001 June. These data are fully described in
 \citet{Cristobal03}. Atmospheric conditions were photometric, with 
seeing ranging from $0.6\arcsec$ to $0.75\arcsec$ FWHM in 2000 April, and 
$0.60\arcsec$-$1.30\arcsec$ FWHM in 2001 June. Source extraction and 
photometry of $K_{\rm s}$-band data was done using SExtractor 
\citep{Bertin96}. Zeropoint uncertainty is 0.03\,mag.
SExtractor BEST magnitudes are provided with errors including 
photon noise. The resulting $K_{\rm s}$ magnitudes are listed
under column (7) of Table \ref{tab:opt-ir}.

\subsection{Redshifts and X-ray Luminosities}\label{sec:z_and_l}

Redshifts, redshift quality flag and the reference source of the redshift are 
shown in columns (8), (9), and (10) of Table 5. The meanings of the redshift 
quality flags are: -1: photometric redshift only, 0: no redshift information,    
1-5: as defined by the DEEP project \citep{deep_lowe} or the CFRS \citep{cfrs3} 
confidence class if the redshift source is CFRS. The redshift of x22
has been solely determined by the H$\alpha$+[NII] complex from the 
Subaru OHS data.  Since this feature is unambiguous, it has been assigned 
the quality flag of 3.

For those with redshift information, we have calculated the X-ray 
luminosity of the source and listed the results in column (13). 
The luminosities have been calculated from the 2-10 keV rest-frame fluxes 
given in Table ~\ref{tab:srcs} and the 
K-corrections have been made using the spectral assumptions described in 
Sect. \ref{sec:xobs}. The luminosities have been corrected for possible 
intrinsic absorption using the same spectral assumption. Because of the 
difference in methods and exact time selections, the luminosities listed here 
can be slightly different from those in the full spectral analysis listed 
in Table \ref{tab:spec}.        
 
\section{Results and Discussion}
\label{sec:disc}

\subsection{X-ray Source Population}

The 23 sources detected in our 80 ks XMM-Newton observation of the
GWS are representative of the X-ray
sources that contribute most to the ``Cosmic X-ray Background'' 
and many of them represent the regime which marks the peak of 
accretion onto SMBHs in centers of galaxies. 

 A dominant population in this field consists of AGNs with
${\rm Log}\;L_{\rm x}\sim$ 44 at $z\sim 1$. Only a few of them
show signs of AGN activity in their optical spectra. Subaru OHS 
NIR spectroscopy of four of the X-ray sources with no previous optical 
signature of AGNs revealed  H$\alpha$+[NII] emission lines showing  
hints of broad H$\alpha$ and/or stronger narrow [NII] lines indicative 
of AGN activity.

\begin{figure}
\epsscale{1.0}
\plotone{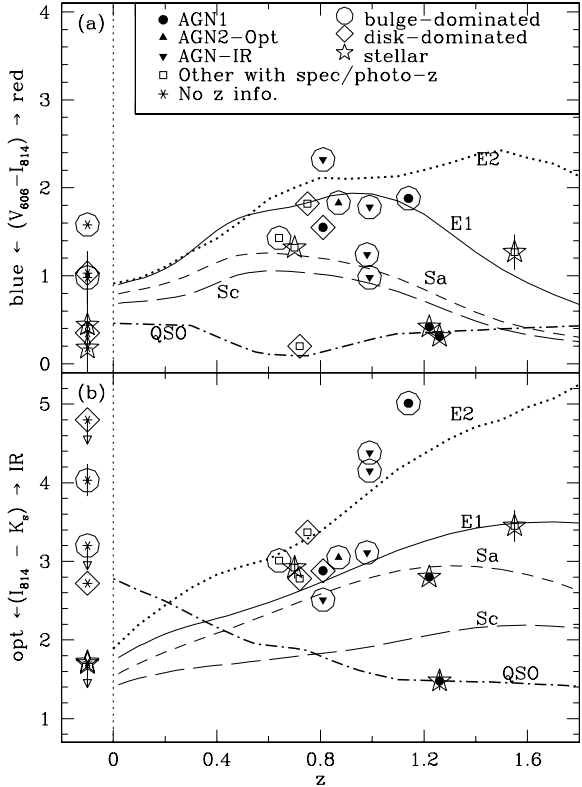}
\caption
 {The $V_{\rm 606}-I_{\rm 814}$ (upper panel) and $I_{\rm 814}-K_{\rm s}$ colors
of the X-ray sources plotted as a function of redshift. Those without
redshift information are plotted left of zero. 
The meanings of the symbols are the same as those in Fig. \ref{fig:hr}. 
One sigma error bars are shown if $\sigma \geq 0.1$.  The solid, short-dashed, 
long-dashed curves are galaxy colors as functions of $z$, 
based on model spectra for Elliptical (E1), Sa, and Sc galaxies after K- and 
evolution corrections given by \citet{poggianti97}. The thick dotted line show 
the colors of a passively evolved  elliptical model by \citet{ka97}. 
The dot-dashed lines are the colors calculated from the mean radio-quiet QSO 
spectrum by  \citet{elvis94}. 
 }
\label{fig:col}
\end{figure}

The host galaxies of the X-ray sources tend to be bulge-dominated and 
four are extremely-red objects (EROs) (or Very Red Objects; VROs) 
($I_{814}-K_{\rm s}\geq 4$). { Also one object (x130) has an upper limit
$I_{814}-K_{\rm s}< 4.8$, which is consistent with being an ERO.} 
Fig. \ref{fig:col} shows $V_{\rm 606}-I_{\rm 814}$
and $I_{\rm 814} - K_{\rm s}$ colors of the X-ray sources 
in our sample as a function of redshift. We have excluded 
x33 (no $K_{\rm s}$ and $V_{\rm 606}$ photometry available),x46 (X-ray source
is off-nucleus, see below), and x83 (Galactic star) from the plot.  Those 
without redshift information are plotted left of z=0.  For reference, 
we plot K- and evolution-corrected galaxy colors for elliptical (labeled 
as E1 in  Fig. \ref{fig:col}), Sa and Sc galaxies from \citet{poggianti97}.  
Model E1 corresponds to Poggianti's model ``E'', which has a star-formation
rate with e-folding time of 1 Gyr. We have neglected the difference 
of cosmological parameters used by us and\citet{poggianti97}, which 
corresponds to a $\sim 10\%$ difference in the age of the Universe.
In plotting these color tracks, we have converted the $I$ and $V$ magnitudes 
to $I_{\rm 814}$ and $V_{\rm 606}$ using the formulae given by \citet{cow99}.  
We also plot the colors of an elliptical galaxy model, which we have 
calculated for our filters and cosmology from the evolving SEDs by 
\citet{ka97} (labeled as E2). The plotted model is for a passively evolving 
galaxy after a short burst at the formation epoch of $z_{\rm f}=4.62$ 
and $L=0.1 L_*$ at $z=0$.  The difference between $K$ and $K_{\rm s}$ magnitudes 
have been neglected, following \citet{Cristobal03}. 


 Fig. \ref{fig:col} shows that the { $I_{\rm 814} - K_{\rm s}$ color} 
of the X-ray sources with bulge-dominated hosts indeed traces that 
of elliptical galaxies 
and, in particular, those of EROs with redshifts are roughly consistent 
with the passively evolving elliptical model (E2). { The 
$V_{\rm 606} - I_{\rm 814}$ color, which is more sensitive to the 
contaminations from star formation and AGN activities, shows a scatter
towards bluer colors from the elliptical regime.} 
The reddest one (x20), with $I_{\rm 814}-K_{\rm s}=5.0$ may be contributed by 
a dust enshrouded AGN or a starburst. Point-like 
sources tend to be distributed towards bluer (QSO) colors, although the 
scatter is large. The scatter may be 
contributed to by unresolved host galaxies, intrinsic scatter in QSO colors 
and/or reddening by intrinsic dust absorption. 
    
\subsection{Bulge Mass and X-ray Luminosity}

The HST imaging of this field has allowed us to find
a significant population of X-ray sources at $z\sim 1$ whose counterparts 
have a resolved bulge component, either as part of a disk+bulge structure
or a pure bulge.  In view of the relationship 
found between bulge mass and the central blackhole mass in nearby galaxies 
\citep{magorrian,merritt}, it is 
interesting to make a first-order estimation of blackhole mass 
($M_{\bullet}$) from the bulge-component of the host galaxy
and compare it with the X-ray luminosity. In the rough estimation below, we take 
the approach of \citet{aller} by first converting the bulge luminosity to
the central velocity dispersion ($\sigma$) of the bulge stellar component using
an empirical relation. Then we use the tight $\sigma - M_\bullet$ 
\citep{merritt,tremaine} correlation to obtain the estimated blackhole mass. 
We use the F814W K-correction for the 
E galaxy in Fig. 18(d) of \citet{fukugita} to calculated the F814W absolute 
magnitude ($M_{\rm F814W}$).  We also assume an early-type galaxy color of  
$$
M_{\rm B_T}-M_{\rm F814W} = 2.1
$$
\citep{fukugita,gonzalez}. We then use the relations, 
\begin{eqnarray}
-M_{\rm B_T}+5\log h_{\rm 70} = 20.5 + 7.7(\log \sigma - 2.3),\nonumber\\
M_\bullet/M_\sun = 1.48\; 10^8\;(\sigma/200)^{4.65},\;\;\; \label{eq:mass}
\end{eqnarray}
\citep{gonzalez,merritt} to obtain the estimated blackhole mass. Because
the F814W band corresponds to the $B$ band at the source rest 
frame of $z\sim 1$, the combination of the K-correction and the magnitude
conversion  to $M_{\rm B_T}$ is insensitive to the assumed galaxy spectral 
energy distribution. 


\begin{figure}
\plotone{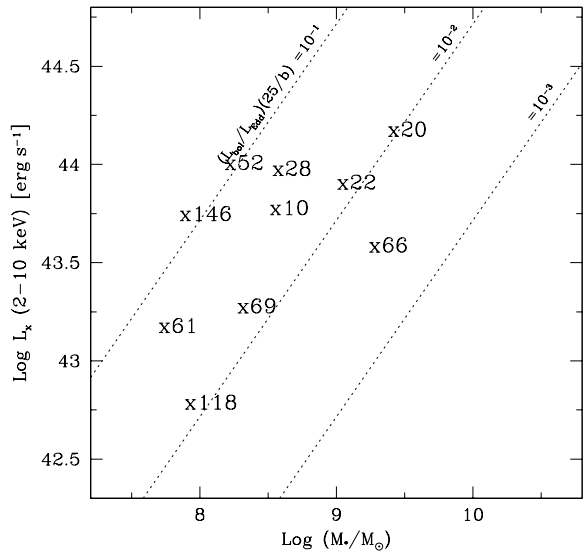}
\caption
 {The X-ray luminosity (absorption-corrected) plotted
 as a function of the central blackhole mass ($M_\bullet$) estimated
 from the bulge luminosity (see text) for the 10 X-ray sources in the
 sample with resolved bulge components. The data points are
 shown with X-name labels. Three lines correspond to 
 $(L_{\rm bol}/L_{\rm Edd})(25/b)=10^{-1},10^{-2}$, and $10^{-3}$, 
 where $L_{\rm bol}=b\,L_{\rm x}$ is the bolometric luminosity of 
 the AGN component and $L_{\rm Edd}$ is the Eddington luminosity 
 corresponding to $M_\bullet$.
}
\label{fig:mbhvslx}
\end{figure}

 Fig. \ref{fig:mbhvslx} shows the estimated $M_\bullet$ versus $L_{\rm x}$
(absorption corrected, see Sect. \ref{sec:z_and_l}) for 10 X-ray sources
in the sample which have a resolved bulge component in the HST WFPC2
F814W image (MDS). 

 It is interesting to estimate the Eddington ratio $L_{\rm bol}/L_{\rm Edd}$
for these X-ray sources, where $L_{\rm bol}$ is the bolometric luminosity of 
the AGN and $L_{\rm Edd}$ is the Eddington luminosity corresponding to the 
blackhole mass. Writing  $L_{\rm bol}=b\;L_{\rm x}$, where $b\approx 25$ 
\citep{elvis94}, we overplot three lines showing  
$(L_{\rm bol}/L_{\rm Edd})(25/b)=10^{-1},10^{-2},$ and $10^{-3}$ in 
Fig. \ref{fig:mbhvslx}. Fig. \ref{fig:mbhvslx} shows that the estimated 
blackhole masses for the 10 AGNs range from  $10^7-10^{10}$ $M_\sun$ and the 
Eddington ratios from 0.3\%-10\%. 

These results have interesting implications on how AGN evolve and the current
stage of evolution for the AGN represented here ($z\sim 1$,
${\rm Log}\,L_{\rm x}\sim 44$).  This is indeed a characteristic 
redshift and luminosity marking the peak of the accretion history 
of the universe.  The result that
these AGN are typically radiating at a few percent may have important
implications on the accretion history and formation of SMBHs. If this 
is a typical Eddington ratio throughout the AGN phase of these
objects, the growth of the blackhole occurs on
a timescale of $\sim t_{\rm s}/(L_{\rm bol}/L_{\rm Edd}) \sim$ 
a few $\times \;10^9$ yrs, where 
$t_{\rm s}\sim 5\;10^7 (\frac{\epsilon}{1-\epsilon})(\frac{0.1}{1-0.1})^{-1}$ 
yrs is the Salpeter timescale, that is the timescale at which the
mass of an object accreting at the Eddington Luminosity grows by 
a factor of $e$ (for a radiative efficiency $\epsilon\sim 0.1$ of 
a standard accretion disk model). This scenario has difficulty in that 
the timescale of a few $\;10^9$ yrs may be too long, while the number 
density of AGN at ${\rm Log} L_{\rm x}\lesssim 44$ decreases beyond 
$z\sim 1$ \citep{ueda1}. Alternatively, it is possible that these AGN have 
gone through a brief luminous phase in the past with 
near-Eddington accretion rates. These may have been observed as more luminous QSOs 
(${\rm Log}\;L_{\rm x}\gtrsim 45$) at $z>2$. Yet another more
exotic possibility is that they are just accreting with a low 
radiative efficiency ($\epsilon << 0.1$), allowing much less time for 
the SMBH to grow.

 We note, however, that there are a number of caveats in interpreting 
these results and drawing conclusions relating the X-ray AGN evolution 
and growth of the SMBH. Firstly, the current MDS database shows the analysis 
for stellar (point-like) images or galaxies (pure bulge, pure disk or 
bulge+disk decomposition), but only limited analysis \citep{vicki99} has been completed for 
decomposing point-like (stellar) nuclei from the host galaxy (stellar+bulge+disk, 
stellar+bulge or stellar+disk). Thus we select against those with strong
AGN components (or with large  $(L_{\rm bol}/L_{\rm Edd}$), which are 
likely to be listed as ``stellar'' in the MDS database or the bulge luminosity
in the database may be contaminated by the central AGN component. 
This situation should be improved in the future, where the HST images
are analyzed with point-like nucleus+host galaxy decomposition.  
Extending this study to other deep fields with X-ray and HST (WFPC2 as
well as ACS) data, including the Extended Chandra Deep Field-South (E-CDFS)
and the COSMOS field, will be a next logical step. Secondly, a much more 
fundamental limitation is that we have assumed the local relations in 
Eq. \ref{eq:mass} are still valid at $z\sim 1$. This assumption is not 
guaranteed to be valid. 

\subsection{Comments on Selected Individual Objects}
\label{sec:indi}

\begin{description}
\item[x8:] This is a typical type 1 QSO at z=1.22 with a broad Mg II line.
\item[x10:] The DEEP optical spectrum shows many narrow emission lines including
  high-excitation lines like [NeV]$\lambda\lambda 3346,3426$, which 
  are unambiguous indicators of AGN activity. Permitted lines (Mg II, H$\beta$)
  are also narrow. Since H$\beta/$/[OIII]$\lambda 5007 \approx 0.7 > 1/3$, these lines
  are not dominated by a Seyfert 2. It is either a Seyfert 2 whose emission lines
  are heavily  contaminated by starburst activity or a narrow-line Seyfert 1 
  galaxy (NLS1) \citep{oster_pogge95}, where H$\beta$ is contributed to from the 
  (narrow end of the) broad line region. Because our X-ray 
  spectral analysis shows no X-ray absorption ($N_{\rm H}<10^{21.5}{\rm cm^{-2}}$; 
  see Table \ref{tab:spec}), the NLS1 interpretation is more plausible.
\item[x11:] The optical counterpart has a very bright stellar (point-like) morphology
  and there is no optical spectrum available for this source to discriminate between 
  a galactic star and a QSO.  However, its X-ray spectrum is inconsistent with a 
  thermal plasma, having a significant residual in the soft part. Also its X-ray
  to optical flux ratio ${\rm Log}\;(f_{\rm x}/f_{\rm R})\approx -0.7$ ($f_{\rm x}$ 
  is measured in 2-10 keV) is well within the AGN regime (roughly between
  -1 and 1; see e.g. \citealt{horn01}). Thus it is most likely to be a QSO. 
\item[x20:] Thanks to the Chandra position, we can identify the X-ray source
  with the brightest bulge-dominated galaxy at $z=1.148$, among four candidates
  apparently interacting with one another indicated by tidal bridges 
  (See Fig. \ref{fig:poststamp}). It is an interesting case where galaxy interactions 
  are possibly feeding the AGN activity at this early stage of the universe. It has a 
  QSO luminosity (${\rm Log} L_{\rm x}=44.2$), but the optical image is
  dominated by bulge component of the host galaxy.  The DEEP spectrum
  shows a broad Mg II line.  The X-ray spectrum shows an absorption of
  ${\rm Log}\;N_{\rm H}\sim 22.3$. This is an example of optical type-1
  X-ray type 2 AGN. This is also an ERO ($I_{\rm 814}-K_{\rm s}=5.0$) and 
  a sub-mm source detected in a deep SCUBA survey \citep{waskett}.
\item[x22:] No previous optical/IR spectroscopic observations existed for this source. 
  Our Subaru OHS/CISCO observation detected H$\alpha$ and $[NII]\lambda\lambda$ 
  6548,6583 emission lines, giving z=0.983. Based on the 
  sign of broad H$\alpha$ and strong $N{II}$, high luminosity 
  ${\rm Log}\;L_{\rm x}\sim 44.2$. While we mark it as an AGN-IR, it
  may well be a type 1 AGN.
\item[x28:] This source is the most conspicuous hard X-ray source in the field with an
  intrinsic absorption of ${\rm Log}\;N_{\rm H}\sim 22.5$ [cm$^{-2}$].  It is a  
  bulge-dominated galaxy with no optical/IR spectroscopy and 
  a photometric redshift of $z=0.76$ \citep{im} based on its V--I color alone. 
  This is most likely a Seyfert 2 based on the X-ray properties.
\item[x46:] The position of the source has been determined with Chandra and the
 relative alignment of the WFPC2 image and X-ray sources has been 
 achieved the three other CXO sources in the field. The X-ray source counterpart is identified 
 with a hot spot just off the patchy irregular starforming galaxy DEEP gss 074\_2638 
 (MDS u2ay1:0019) at z=0.432. If we assume that the X-ray source is at the same 
 redshift as this irregular galaxy, the luminosity would be 
 ${\rm Log}\;L_{\rm x}\approx 42.8$. 
 This is too luminous for an ultraluminous X-ray source (ULX).  We also note 
 that there is a nearby edge-on 
 disk galaxy (DEEP gss 074\_2237, z=0.156). The X-ray source is 4$\arcsec$ away 
 (projected distance of 10 kpc) from its nucleus and located towards the direction 
 perpendicular to the disk.  Therefore the  X-ray source could be a ULX 
 associated with the halo of the galaxy. However, even if the X-ray source was at the 
 redshift of this disk galaxy, its luminosity would still be 
 ${\rm Log}\;L_{\rm x}= 41.8$, again well above the ULX regime. One possibility is 
 that this irregular galaxy is undergoing a merging process and the X-ray source is 
 at the nucleus of one of the merging galaxies, or it may simply be a background QSO. 
\item[x52:] The DEEP spectrum show no broad lines and the location
   of MgII shows only absorption.  Other features include [OII]$\lambda 3727$ 
   and [NeIII]$\lambda 3869$. The presence of [NeV]$\lambda 3426$ is suggested but 
   uncertain. We observed this object with
   Subaru OHS and found a moderately broad H$\alpha$ 
   (FWHM $\sim 2000 {\rm km\,s^{-1}}$) and a strong [NII] 
   doublet. (Fig. \ref{fig:ohs}). { The fact that there is no broad MgII line
   but a moderately broad  H$\alpha$ suggests that the AGN is obscured by a dust
   cloud. This is consistent with the hard color of this object, with the second 
   largest HR(2-4.5 keV/0.5-2 keV) in the sample. See Fig. \ref{fig:hr}.}  
   This is an ERO ($I_{\rm 814}-K_{\rm s}=4.4$). 
\item[x66:] We find no indication of an AGN in the DEEP spectrum. 
   Like x52, our Subaru OHS observation revealed possible AGN activity
   through the detection of a moderately broad $H\alpha$ (FWHM $\sim 2000 {\rm km\,s^{-1}}$)
   line and a strong [NII] doublet. { This is an obscured AGN similar to x52 and
   has a hardest HR(2-4.5 keV/0.5-2 keV).} There is also a hint of [NeIII]$\lambda 3869$, 
   a line which is often stronger in AGN than starforming galaxies.   
\item[x69:] A spectroscopic redshift of $z=0.995$ has been determined 
  by our Subaru observation in close agreement with the photometric redshift determined 
  by \citet{brunner} of $z_{\rm ph}=0.935$.   
  The Subaru OHS spectrum seems to indicate either strong [NII]$\lambda 6583$ or 
  broad H$\alpha$ suggestive of AGN activity. This is an ERO 
  ($I_{\rm 814}-K_{\rm s}=4.1$).
\item[x83:] Because of the low X-ray to optical flux ratio 
   (${\rm Log}\;(f_{\rm x}/f_{\rm R})\approx -3$) for this bright 
   optical object (F606W=14.8), this is certainly a Galactic 
   star.    
\item[x125:] Although the MDS database shows that it is a point 
   source, its color is consistent with an elliptical galaxy at
   the photometric redshift of $z_{\rm ph}=1.55$ \citep{brunner}. 
   The optical counterpart is probably dominated by the host 
   galaxy.
\item[x146:] This object is detected only in the hard (2-8 keV)
   X-ray band. The DEEP optical spectrum clearly shows 
   [NeV]$\lambda 3426$ emission and strong [NeIII]$\lambda 3869$. 
   This is a typical Seyfert 2 galaxy with absorbed X-ray spectrum. 
\end{description}

\section{Summary}

\label{sec:sum}
\begin{enumerate}
\item We summarize the properties of X-ray sources in the GWS detected with
 XMM-Newton and discuss their morphology, optical and NIR spectra.  
\item Ten out of the 23 X-ray sources have spectroscopic information
 available in the optical band through the DEEP and CFRS projects.
 Four of them have broad emission lines and thus can be classified 
 as type 1 AGN/QSOs. Two others have high excitation lines indicative
 of narrow line region of active galactic nuclei. Others do not reveal
 AGN signatures in their optical spectra.
\item We have obtained NIR spectra for five of the sources 
 which did not reveal AGN signatures in their optical spectra using the
 CISCO/OHS spectrograph on the Subaru Telescope.  In four of the five,
 we detect broad H$\alpha$ and/or strong [NII] lines suggestive
 of AGN activity.
\item { The X-ray hardness ratios and spectroscopic analysis show
  that a major part of the observed X-ray sources consists of 
  a mixture of absorbed and unabsorbed AGNs. One of these AGNs 
  (x20) has an absorbed X-ray spectrum with a broad MgII emission line 
  and is another example of an optical type 1/X-ray type 2 AGN.}
\item The $V_{\rm 606}-I_{\rm 814}$ and $I_{\rm 814}-K_{\rm s}$ colors
  of the X-ray sources with bulge-dominated hosts well trace those of
  evolving elliptical galaxies. Sources with stellar morphologies show
  bluer colors with some scatter, and consistent with AGN(QSO) with
  a host galaxy contamination and/or dust absorption. 
\item Ten X-ray sources with redshift information have a resolved bulge. Based
 on a number of assumptions concerning the bulge luminosity-blackhole mass relation,
 we estimate that these AGN have luminosities ranging from 0.3-10\%
 of the Eddington luminosity, suggesting that these massive blackholes
 have previously gone through massive accretion (maybe as luminous QSOs) or 
 that their lifetimes are relatively long ($\gtrsim 10^9$ yrs).        
\end{enumerate} 

\acknowledgments
 
 The authors are indebted to the efforts of the DEEP team for their
observations, analysis, and archiving. This work has been supported 
by the NASA Grant NAG5-10875 to TM (Long-Term Space Astrophysics), 
NASA Grant NAG5-3651 to REG (XMM-Newton Mission Scientist support).

\end{document}